\documentclass[reprint,showpacs,amsmath,amssymb,nofootinbib]{revtex4-1}
\usepackage{graphicx}
\usepackage{dcolumn}
\usepackage{bm}
\usepackage{natbib}

\renewcommand{\a}{\approx}
\newcommand{\p}{\partial}

\renewcommand{\o}{\Box_M}
\renewcommand{\sc}{\scriptscriptstyle}
\newcommand{\h}{\phi_{\sc (2)}}
\renewcommand{\l}{\phi_{\sc (4)}}
\newcommand{\bh}{\bar{\phi}_{\sc (2)}}
\newcommand{\be}{\begin{equation}}
\newcommand{\ee}{\end{equation}}	

\newcommand{\R}{\rho}
\newcommand{\n}{\lambda}

\newcommand{\w}{\omega}

\begin{document}

\title{More about scalar gravity}

\author{E. Bittencourt$^1$} \email{bittencourt@unifei.edu.br}
\author{U. Moschella$^2$} \email{Ugo.Moschella@uninsubria.it}
\author{M. Novello$^3$}\email{novello@cbpf.br}
\author{J.D. Toniato$^3$} \email{toniato@cbpf.br}

\affiliation{$^1$Instituto de Matem\'atica e Computa\c c\~ao, Universidade Federal de Itajub\'a, Itajub\'a, MG 37500-903, Brazil}
\affiliation{$^2$Universit\`a degli Studi dell'Insubria - Dipartimento DiSAT\\
Via Valleggio 11 - 22100 Como - Italy and\\
INFN, Sez di Milano, Via Celoria 16, 20146, Milano - Italy}
\affiliation{$^3$Instituto de Cosmologia Relatividade Astrofisica ICRA - CBPF\\
Rua Dr. Xavier Sigaud 150 - 22290-180 Rio de Janeiro - Brazil\\}

\date{\today}

\begin{abstract}
We discuss a class of models for gravity based on a scalar field. 
The models include and generalize the old approach by Nordstr\"om
which predated and in some way inspired  General Relativity. 
The class include also  a model that we have recently introduced and discussed in its cosmological aspects (GSG).
We present here a complete characterisation of the Schwarschild geometry as a vacuum solution of GSG and sketch a discussion  of the
first Post-Newtonian approximation.
\end{abstract}


\maketitle

\begin{sloppypar}

\section{Introduction}
One hundred years have passed and General  Relativity (GR) is still the paradigm for scientific thinking about gravity.
General Relativity is probably the most beautiful in the realm of physical theories and one of the greatest achievements of humanity \cite{time}. More than that, even the man in the street has heard
about  space-time and its curvature and how Einstein discovered all that.
Nowadays this worldwide excitation is renewed after the claim that gravitational waves have been observed resulting from the coalescence of two black holes. The fine details of the waveform have been predicted starting from 1999 in a brave tour de force calculation \cite{damour1,damour2, damour3,damour4,damournagar,damournagar1,damournagar2} exploiting all the highly subtle features of general relativity and their agreement with the experimental outcome is astonishing.

Nonetheless,  even just an inattentive look to contemporary theoretical and experimental researches in gravity  \cite{wiki}  and particle physics
shows
a luxuriant vegetation of alternative models and theories  where the need to go beyond GR is claimed as a necessity.
Indeed, although no experimental fact has yet invalidated the general theory of relativity and, on the contrary, GR agrees with extremely high precision to local observational tests both for weak and strong gravity \cite{will}, the anomalies observed at the astrophysical or cosmological level are kind of embarrassing. Among them, the most important are the abnormal dynamics of galaxies and the recently discovered acceleration of the expansion of the Universe. Dark Matter and Dark Energy are invoked to rescue and care of GR but it may very well be possible that the theory simply does not describe correctly gravity at those scales.

Still, most alternative models confirm the paradigmatic role of GR in that they are adaptations of it with more or less important
modifications which however show how difficult is to deform GR by leaving untouched its successes and beauty. One may also wonder whether, from a conceptual viewpoint, it is more expensive to add dark energy and dark matter to the energy-momentum tensor of the cosmic fluid rather than producing highly speculative modifications of GR such as imagining that the visible universe stays on a brane which has a companion brane somewhere else or that the action of the gravitational field is $f(R)$ rather than simply the Ricci scalar $R$, where $f$ is a function having a bunch of ad hoc properties and parameters that render the theory more flexible (but also, perhaps, uglier).

Here we invoke   the well known Feyerabend's epistemological
{\em ``{counterrule} that urges us to develop hypotheses inconsistent with well-established facts.... The advice (which goes back to Newton and which is still very popular today) to use alternatives only when refutations have already discredited the orthodox theory puts the cart before the horse. Also, some of the most important formal properties of a theory are found by contrast, and not by analysis. A scientist who wishes to maximize the empirical content of the views he holds and who wants to understand them as clearly as he possibly can must therefore introduce other views; that is, he must adopt a pluralistic methodology. He must compare ideas with other ideas rather than with `experience' and he must try to improve rather than discard the views that have failed in the competition."} \cite{feyer}.

One view which has very early failed in the competition \cite{nord1,nord2, ravndal, norton} is the simple naive idea that the gravitational degrees of freedom of the world may be encoded in a single scalar field -- the relativistic generalization of the Newtonian potential \cite{nord1}.
Following Feyerabend's counterrule, in two recent papers \cite{JCAP,gsgcosmo} we went back to that old failed idea and tried to improve it by introducing and beginning to explore a toy model of gravity based on a single scalar field which in a way overcomes certain difficulties of the old times. Here we continue that  investigation to see how far we can go.

\section{Nordstr\"om's  theories of gravity}

Let us begin by a brief summary of  the key features of Nordstr\"om's theories, referring the reader to \cite{ravndal,norton,bruneton,der1,der2} for a detailed account. 

Nordstr\"om's original idea is that gravity is mediated by a single massless scalar field $\phi$ in Minkowski spacetime. Two geometries enter in the construction of the theory: the metric  $\eta_{\mu\nu}$ of the Minkowski spacetime and the "physical" metric $g_{\mu\nu}$ 
where the  matter's dynamics takes place. The founding hypothesis  is that $g_{\mu\nu}$ is conformal to the Minkowski metric:
\be
g_{\mu\nu} = a^2(\phi) \eta_{\mu \nu}\,, \label{conformal}
\ee
the conformal factor  $a(\phi)$ being a given function of the scalar field $\phi$;  the choice of $a(\phi)$ characterizes one particular Nordstr\"om's theory.
 All the above said, the action for gravity plus matter may be written as follows:
\begin{eqnarray}
&& S= S_{gravity}(\phi) + S_{\text {matter}}(\psi,g_{\mu\nu}) \label{nord}\,,
\\[2ex]
&& S_{gravity}(\phi) =  \frac{1}{\kappa c} \int \sqrt{ -\eta} \,\eta^{\mu\nu}\, \partial_\mu\phi\,\partial_\nu\phi \, {d^4x}\,, \\[2ex]
&& S_{matter}(\psi,g_{\mu\nu})=\frac{1}{c}\int \sqrt{-g}\,L_m\,d^4x\,,
 \end{eqnarray}
where $\psi$ denotes globally all non-gravitational fields, $g$ is the determinant of $g_{\mu\nu}$ and $\kappa =  {8\pi G}/c^4$.  In our conventions \cite{adler}  the Minkowski metric written in Cartesian coordinates is mostly minus: $\eta_{\mu\nu}= diag(1,-1,-1,-1)$, but it can be written in any coordinate system.

The hypothesis that the matter fields are minimally coupled to the physical metric $g_{\mu\nu}$ warrants the validity of the weak equivalence principle.
The action (\ref{nord}) leads to the following field equations:
\begin{eqnarray}
&&\Box_M \phi = \frac{1}{\sqrt{-\eta}}\partial_\mu (\eta^{\mu\nu} {\sqrt{-\eta}}\partial_\nu \phi )=  -\frac{4\pi G}{c^4} a' a^3\, T, \label{fieldeq} \\[2ex]
&&T^{\mu\nu}_{\ \ \ ;\nu} = 0\,,\label{2}
\end{eqnarray}
with $a'=da/d\phi$\,. Here the energy-momentum tensor of the non-gravitational fields is defined as the variational derivative w.r.t. the physical metric $g_{\mu\nu}$
:
\be
\label{tmunu_q}
T_{\mu\nu} = \frac{2}{ \sqrt{ -g}} \frac{\delta (\sqrt{-g}L_m)}{\delta g^{\mu\nu}}\, ;
\ee
the covariant derivative in Eq. (\ref{2}) is also taken w.r.t. to $g_{\mu\nu}$ and  the trace is given by $T = T^{\mu\nu}g_{\mu\nu}$\,.

The scalar curvature of the conformal geometry (\ref{conformal}) reads
\be
R =  \frac{6\, \Box_M a( \phi )}{a^3(\phi)}. 
 \label{ricci}
\ee
By using this relation, Equation (\ref{fieldeq}) may be rewritten as follows
\begin{eqnarray}
R &=& - \ \frac{24\pi G}{c^4} (a'(\phi))^2 T + \frac{6 a''(\phi) }{a(\phi)} g^{\mu\nu}\partial_\mu \phi\partial_\nu\phi \,. \label{fieldeq2}
\end{eqnarray}
Nordstr\"om originally introduced two models: a first model \cite{nord1} that immediately failed and a second model \cite{nord2} that fixed some of the shortcomings of the first  one and corresponds to the choice $a(\phi) = \phi$. In this case   Eq. (\ref{fieldeq2}) reduces to
\begin{eqnarray}
R &=& - \ \frac{24\pi G}{c^4}  T   \label{fieldeq3}
\end{eqnarray}
(the minus sign at the RHS is because of our conventions on the curvature tensor \cite{adler}). This reformulation is due to  Einstein and Fokker \cite{fokker}. It is the first
purely geometric description of gravity and, together with GR,  the only theory  satisfying the Strong Equivalence Principle (see e.g. \cite{will,der2}). 

How deeply this Equation (\ref{fieldeq3}) influenced Einstein's path to General Relativity? This question will probably stay without an answer.

The variation  of $S_{\text {matter}}$ w.r.t to the Minkowski metric gives  the ``Einstein-frame" stress-energy tensor
\be
\label{tmunu_mink}
\widetilde{T}^{\mu\nu} = \frac{2}{ \sqrt{ -\eta}} \frac{\delta (a^4\sqrt{-\eta}\,L_m)}{\delta \eta_{\mu\nu}} = {a^6}\,T^{\mu\nu}\,.
\ee
By introducing an analogous Einstein-frame stress-energy tensor for the scalar field $\tilde t^{\mu\nu}$, translation invariance w.r.t. the inertial coordinates gives the conserved current
\be
\label{cons_tmunu_mink}
\partial_{\nu}(\widetilde{T}^{\mu\nu}+ \tilde t^{\mu\nu})=0\,.
\ee

\section{Geometric scalar  theories of gravity}
One major shortcoming of Nordstr\"om's theories of gravity is that conformal invariance forbids the minimal coupling of the electromagnetic field to the scalar field $\phi$, irrespectively of the choice of the conformal factor $a(\phi)$ in the action (\ref{nord}).   Thus, Nordstr\"om's theories altogether imply that gravity does not deflect light and they are ruled out by the astronomic observations.

However, the conformal transformation (\ref{conformal})  is nothing but the simplest way to relate the physical metric to the Minkowski metric. 
Might that relation be generalized? This question has been raised and answered by Bekenstein \cite{beke} within the two-geometries paradigm for gravity. In this context, one metric describes that bare gravitational field while the other defines the  geometry 
which is physically seen by matter.  Bekenstein found that, by assuming the validity of causality and the weak equivalence principle (and a few other simplicity and minimality criteria),
the most general relation that can be established between the "gravitational" metric $\tilde g_{\mu \nu} $ and the "physical" metric $g_{\mu \nu} $ is a {\em{disformal}} one:
\begin{eqnarray}
{g}^{\mu\nu} = A(\phi,\omega) \tilde g^{\mu \nu} + B(\phi,\omega) \partial^\mu \phi \partial^\nu \phi\,,
\label{disformal0}
\end{eqnarray}
where $\phi$ is scalar field and $\omega =\tilde g^{\mu\nu} \partial_\mu \phi \partial_\nu \phi$ is the scalar length of the gradient of the field.

In this paper (and in the two preceding ones \cite{JCAP,gsgcosmo})
we take a huge jump backwards by reconsidering models based just on a single scalar field.
As a possible way out of some of the difficulties that Nordstr\"om's theories have because of 
 the conformal prior (\ref{conformal}) we  explore the possibility that the physical metric be related to the Minkowski 
 metric by a disformal transformation. 
This hypothesis immediately permits the minimal coupling of the electromagnetic field  to the physical metric.
Also, we  allow for more general Lagrangians for the gravitational part of the action and in particular we admit modifications of the kinetic term.

Summarizing,  we propose to enlarge the Nordstr\"om's family of theories as follows. 
The founding hypothesis  is that the physical metric $q_{\mu\nu}$ is disformal to the Minkowski metric:
\be
q^{\mu\nu} = A(\phi,\omega) \eta^{\mu \nu} + B(\phi,\omega) \partial^\mu \phi \partial^\nu \phi \label{disformal}
\ee
where $\omega =\eta^{\mu\nu} \partial_\mu \phi \partial_\nu \phi$.The action for gravity plus matter may be written as follows:
\begin{eqnarray}
&& S= S_{\text {gravity}} (\phi) + S_{\text {matter}}(\psi,q_{\mu\nu})\, , \label{disf}\\[2ex]
&&  S_{\text {gravity}} = \frac{1}{\kappa c} \int L(\phi, \partial _\mu\phi) \,  \sqrt{ -\eta}  \,d^4x  \,, \label{disf2}\\[2ex]
&&  S_{matter}(\psi,q_{\mu\nu})=\frac{1}{c}\int \sqrt{-q}\,L_m\,d^4x.
\label{disformalgsg}
\end{eqnarray}
 When $A= a^{-2}(\phi)$, $B = 0$ and $L$ is the Lagrangian of a massless Klein-Gordon field we are back to  Nordstr\"om.\footnote{We thank Nathalie Deruelle for pointing out an imprecision in the identification proposed in Appendix 4 of \cite{JCAP}.}

Theories where the disformal coupling plays a role have attracted some interest only recently \cite{Kaloper:2003yf,libe,hornd,vernizzi,der3,Brax,Zumalacarregui:2013pma,Koivisto:2008ak,Zumalacarregui:2010wj,Noller:2012sv,vandeBruck:2012vq,Zumalacarregui:2012us,vandeBruck:2013yxa,Brax:2013nsa,
Brax:2014vva,Brax:2014zba,Sakstein:2014isa,Sakstein:2014aca,vandeBruck:2015ida,Koivisto:2015mwa,Hagala:2015paa,Domenech:2015hka}.
In particular scalar-tensor theories  may be constructed along the lines of the seminal paper \cite{damour} by modifying the term describing the coupling of matter to the physical metric exactly as in Eq. (\ref{disformal}). The Horndeski class \cite{hornd,libe} and the ``beyond Horndeski" class \cite{vernizzi} are scalar tensor theories of this kind.
The disformal coupling plays a role also in the relativistic theories of MOND \cite{bruneton2}.

However, to the best of our knowledge, the model presented in  \cite{JCAP}  is the first (and up to now the only) one where a scalar field disformally coupled is used
to modify the original Nordstr\"om's idea and describe gravity in the context of {\em purely scalar} theory.

We call a theory belonging to this family a geometric scalar theory of gravity for the obvious reason that matter interacts with gravity only through minimal coupling to the physical metric (\ref{disformal}). A general geometric scalar theory of gravity is characterized by three functions: the functions $A$ and $B$ characterizing the metric and the Lagrangian $L$ of the scalar field. Of course those functions cannot be completely arbitrary; they should at least warrant the Lorentzian character of the metric and the well-posedness of the Cauchy problem for the field equations. We will come back to the above questions elsewhere.

The Sherman-Morrison lemma (or an elementary direct calculation)  implies that the infinite series defining the covariant physical metric may be summed and the result is again a binomial-like metric (but of course the gradient of the field enters in the potentials at the denominator):
\begin{equation}
q_{\mu\nu} = \frac{1}{A} \, \eta_{\mu\nu} - \frac{B}{ A^2+AB\omega} \, \partial_{\mu}\phi \,\partial_{\nu} \phi \label{cova} \,.\end{equation}
The determinant of the matrix $q^{\mu\nu}$ is also computed by the Sherman-Morrison lemma:
\begin{eqnarray}
\det q^{\mu\nu} =  -A^4-A^3  B\omega\,. \end{eqnarray}

As before two stress-energy tensors may be defined by variating the action \eqref{disf} w.r.t. the physical metric,
\begin{equation}
T_{\mu\nu}= \frac{2}{\sqrt{-q}}\,\frac{\delta(\sqrt{-q}\,L_m)}{\delta q^{\mu\nu}}\,,
\end{equation} or w.r.t. the Minkowski metric, as in (\ref{tmunu_mink}); the relation between them is now a little trickier:
\begin{eqnarray}
\widetilde{T}_{\mu\nu}&=&\frac{1}{\sqrt{A^3(A+B\w)}}\left(A\,T_{\mu\nu} \right. + \nonumber \\[1ex]
&&+ \left. \frac{B}{A+B\w}\, q^{\beta\alpha}\,T_{\beta(\mu}\partial_{\nu)}\phi \partial_{\alpha}\phi\right).
\end{eqnarray}

\section{A case study: GSG }

When $T=0$  Nordstr\"om's  theories  coincide with the flat space massless Klein-Gordon theory irrespectively of the conformal factor (see Eq. (\ref{fieldeq})).

A similar -- but also distinct -- feature is shared by a particular class of geometric scalar theories of gravity \cite{JCAP} that in vacuo
reduce to the massless Klein-Gordon equation but now {\em w.r.t. the curved spacetime physical metric $q_{\mu\nu}$}.
This possibility is opened by having considered more general Lagrangians for the scalar field.
The choice  considered in \cite{JCAP} is to multiply the standard kinetic term  by a field dependent amplitude (potential):
\be
L = V(\phi)\,  \eta^{\mu\nu} \partial_\mu \phi \partial_\nu \phi\,.
\ee
The potential may be reabsorbed by a simple field redefinition but it is useful to keep it explicit.

The second hypothesis consists in restricting the disformal metric to the following particular case:
\begin{equation}
\label{4jul}
q^{\mu\nu} = \alpha(\phi) \, \eta^{\mu\nu} + \frac{\beta(\phi)}{\w} \, \partial^{\mu}\phi \,\partial^{\nu} \phi\,,
\end{equation}
where the functions $\alpha$ and $\beta$  depend only on $\phi$ (and do not depend on $\w$). Eq.  (\ref{cova}) now reads
\begin{equation}
q_{\mu\nu} = \frac{1}{\alpha} \, \eta_{\mu\nu} - \frac{\beta}{\alpha
\, (\alpha + \beta) \, \w} \, \partial_{\mu} \phi \, \partial_{\nu} \phi \,.
\label{291}
\end{equation}

With  the above assumptions on the Lagrangian $L$ and the physical metric  $q_{\mu\nu}$,  the field equation in vacuo reduces to the Klein-Gordon equation  relative to the metric $q_{\mu\nu}$
\begin{equation}
\Box \, \phi
= \frac{1}{\sqrt{- q}}
\partial_{\mu} ( \sqrt{- q} \,q^{\mu\nu} \partial_{\nu} \phi 
 )=0  \label{kkgg}
\end{equation}
provided  the following condition holds:
\begin{eqnarray}
 \alpha+\beta=\alpha^3 V\,. \label{pot}
\end{eqnarray}
Contrary to what happens in the Nordstr\"om case, here the field equation keeps its nonlinearity and the gravitational scalar field is self-interacting. It can be linearised by the same field redefinition used to reabsorb the potential, which however requires some care.

In the following we will set $c=G=1$. In \cite{JCAP} attention has been focused on a particular model based on a concrete choice of the functions $\alpha$ and $\beta$ entering in the physical metric (\ref{disformal}). The following conditions have been imposed:
\begin{enumerate}
\item The theory has the correct Newtonian limit.
\item Condition (\ref{pot}) holds.
\item The Schwarzschild geometry is an exact solution of the field equation (\ref{kkgg}).
\end{enumerate}
A solution to the above requirements is provided by the following potentials:
\begin{eqnarray}
&& \alpha = \exp({- 2 \,\phi})\,, \label{pota}\\[2ex]
&& \alpha+\beta=\frac{(\alpha-3)}{4}^2\,, \label{potb} \\[2ex]
&& V(\phi) = \frac{1}{4} \left(e^{\phi} -3 e^{3 \phi}\right)^2\,.\label{13jul2}
\end{eqnarray}

\section{Schwarzschild in GSG}
Here we further dwell on the way the Schwarzschild solution arises in the above particular GSG model. In doing this we will also shed light on the general structure of disformal transformations of a given metric.

Let us start by writing the Minkowski metric in spherical coordinates
\begin{equation}
ds^2_M={dt^2}- d\R^2- \R^2 d\Omega^2.
\end{equation}
\vskip 10pt
The working hypothesis is that the field $\phi =-\frac 12  \log \alpha$  depends only on the radial coordinate $\R$.
The spherical symmetry together with Eqs (\ref{pota}) and (\ref{potb}) imply that the physical metric can be  written as follows:
\begin{equation}
ds^2=\frac{dt^2}{\alpha}-\frac{4d\R^2}{(3-\alpha)^2}-\frac{\R^2}{\alpha}d\Omega^2\,. \label{sphericmetric}
\end{equation}
The non-zero components of the Ricci tensor are
\begin{eqnarray}
R^{0}_{0} &=& -\frac{3(\alpha-3)(\alpha-5)}{16}\frac{\alpha'^2}{\alpha^2} + \nonumber   \\[1ex]
 && + \ \frac{(\alpha-3)^2}{4\alpha}\left(\frac{\alpha'}{\R} + \frac{\alpha''}{2}\right)\,,  \\[2ex]
R^{1}_{1}&=& -\frac{3(\alpha-9)(\alpha-3)}{16}\frac{\alpha'^2}{\alpha^2}- \frac{3(\alpha-3)}{2\R}\frac{\alpha'}{\alpha}+ \nonumber  \\[1ex]
 && + \ \frac{3(\alpha-3)^2}{8} \frac{\alpha''}{\alpha}\,, \\[2ex]
R^{2}_{2} = R^{3}_{3} &=& -\frac{3(\alpha-3)(\alpha-5)}{8}\left( \frac{\alpha'^2}{2\alpha^2} - \frac{\alpha'}{\alpha\R}\right) + \nonumber   \\[1ex]  
 && + \ \frac{(\alpha-3)^2}{8}\frac{\alpha''}{\alpha} -\frac{(\alpha-1)(\alpha-9)}{4\R^2}\,. 
\end{eqnarray}
The GSG field equation  in the above coordinates is explicitly written as follows:
\begin{equation}
(3-\alpha) \partial_\R  \left(\frac{\R^2 (3-\alpha)}{  \alpha^\frac{5}2 }\alpha'\right) = 0\,.
\end{equation}
When $\alpha\not = 3$, the prefactor $(\alpha -3)$ can be removed.  We get
\begin{eqnarray}
\alpha'=-\frac{4M}{\R^2} \frac{\alpha^{5/2}}{(3-\alpha)}\ \label{eq_alpha_split1}\,,
\end{eqnarray}
where $M$ is an integration constant.
A further integration gives
\begin{equation}
 \frac{\alpha-1}{\alpha^{3/2}}-\frac{2 M}{\R}=c\,, \label{qq}
\end{equation}
where $c$ is another integration constant.  
Also
\begin{eqnarray}
\alpha''=\frac{8 M \alpha^{5/2} \left[3 M \alpha^{3/2}(\alpha-5) - \R(
   \alpha^{2}+6 \alpha-9)\right]}{\R^4 (\alpha-3)^3} \label{eq_alpha_split2}\,.\qquad
   \end{eqnarray}
   Inserting the above expressions (\ref{eq_alpha_split1}) and (\ref{eq_alpha_split2}) into the Ricci tensor we get
\begin{eqnarray}
R^{0}_{0} &=& 0\,,\\[2ex]
R^{1}_{1}   &=&  \frac{3 \alpha ^{3} M }{\R^3}  \left (\frac{\alpha-1}{\alpha^{3/2}}-\frac{2 M}{\R}\right )= \frac{3 c\alpha ^{3} M }{\R^3}  \,,      \\[2ex]
R^{2}_{2} =R^{3}_{3} &=&   \frac{ (\alpha -9)  \alpha ^{3/2}}{4
   \R^2}  \left (\frac{\alpha-1}{\alpha^{3/2}}-\frac{2 M}{\R}\right ) \\[1ex]
  &=&\frac{c (\alpha -9)  \alpha ^{3/2}}{4\R^2}\,,
 \end{eqnarray}
and therefore the Ricci tensor vanishes identically if and only if $c=0$.

\subsection{Study of the region $1<\alpha <3$.}
\begin{figure}[h]
\begin{center}
\includegraphics[scale=0.8]{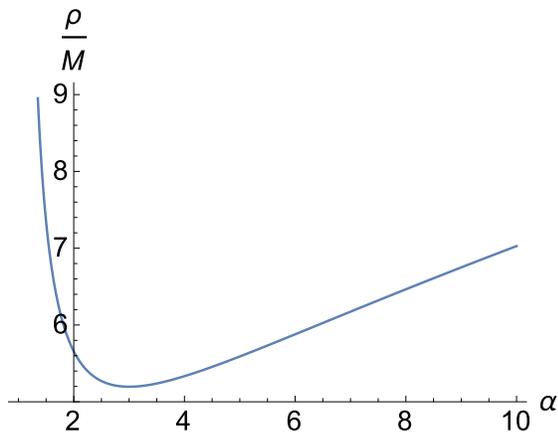}
\end{center}
\caption{Plot of the function  ${\R}/{M}$. }
\label{fig1}
\end{figure}

Let us therefore choose $c=0$. When $1<\alpha<3$, the function
\begin{equation}
 \frac{\R}{M}  = \frac{2 \alpha ^{3/2}} { (\alpha -1)}\,, \label{qq2}
\end{equation}
is one to one (see Eq. (\ref{eq_alpha_split1}) and Fig. \ref{fig1}) and can be inverted (which amounts to solving a third degree algebraic equation). The domain of the inverse  is the region $3 \sqrt 3 M < \R < \infty$.

By inserting the so-constructed function $\alpha(\R)$ in Eq. (\ref{sphericmetric}) we find not only a spherical symmetric solution of the GSG equation but also a spherical symmetric solution of the Einstein's equations in vacuo, since the Ricci tensor vanishes.
By Birkhoff's theorem it must be a portion of the Schwarzschild solution in disguise.
To see it let us introduce a new radial variable as follows:
\begin{equation}
r =  \frac{\R}{\sqrt {\alpha(\R)}}  = \frac{2M\alpha(\R) } { \alpha(\R) -1}\,, \label{qq2bis}
\end{equation}
so that
\begin{equation}
\alpha=\frac 1{1-\frac{2 M}{r}}\,, \ \ \ \ \mbox{and} \ \ \  \R =  \frac{r}{\sqrt {1-\frac{2 M}{r}}}\,.
\end{equation}
Since
\begin{equation}
\frac{\partial \R }{\partial r}=\frac{r-3 M}{r-2 M} \frac 1{\sqrt{1-\frac{2 M}{r}} }\,,
\end{equation}
the coordinate change is well-defined in the whole domain of definition of the function $\alpha(\R)$, i.e. the half-line $3 \sqrt 3 M < \R < \infty$, which is mapped onto the half-line $r>3M$. It is now easy to see that  in that region
the line element becomes \begin{equation}
ds^2=\left(1-\frac{2 M}{r}\right){dt^2}-\frac{dr^2}{1-\frac{2 M}{r}}-{r^2}d\Omega^2\,. \label{sphericmetricschw}
\end{equation}
This result was sketched in \cite{JCAP}. 

Consider now the solution (\ref{sphericmetricschw})
in the whole range $0<r<\infty$.
It is immediate to verify that
\begin{equation}
\Box \log \alpha =
 \frac{1} {r^2} \p_{r} \left[r^2\left(1-\frac {2M}r\right) \partial_r \log\frac1{1-\frac {2M}{r} }\right]=0\,.  \label{gsg2}
 \end{equation}
However, this globally extended Schwarzschild solution of the GSG-like equation (\ref{gsg2}) cannot be written {\em globally}
as a disformal transformation of the Minkowski metric, as in Eq. (\ref{4jul}), in the whole domain  $0<r<\infty$. Still, this can be done piecewise in a rather sophisticated  way which we are going to describe now.

\subsection{Radial geodesics. The region $3<\alpha$.}
To study what happens at the surface $\R= 3 \sqrt 3 M $ (aka $r=3M$) let us briefly examine the radial timelike geodesics.
The geodesic equation relative to the time coordinate $t$ can be integrated as usual \cite{adler}:
\begin{eqnarray}
&&\frac{d^2 t}{ds^2 }  = \frac{\dot t \dot \R  \alpha '(\R)}{\alpha (\R)} \quad\Longrightarrow\quad \dot t =\lambda\alpha. \label{geo_1}
\end{eqnarray}
We set $\lambda=1$ so that at infinity (i.e. $\alpha =1$) proper time and coordinate time coincide. The radial equation becomes
\begin{eqnarray}
&&\dot{\R}^2=\frac{(3-\alpha)^2}{4}\left(\frac{\dot{t}^2}{\alpha}-1\right) = \frac{(3-\alpha)^2}{4}\left({\alpha}-1\right)\,. 
\end{eqnarray}
All test particles reach the surface $\R=3\sqrt{3}M$ with proper velocity zero. The radial acceleration
\[
\ddot \R = \frac{\alpha^{3/2} (3 \alpha-5) M}{2 \R^2} = \frac{(\alpha-1)(3 \alpha-5)}{4 \R}\,,
\]
becomes positive when $\R$ is smaller than a certain critical value $\R_0$ corresponding to $\alpha( \R_0 )=5/3$.
This result looks strange; it seems to indicate that radial trajectories start to slow down and stop at $\R = 3\sqrt 3 M$. But the geometry is nothing but the Schwarzschild geometry and we know in advance that nothing special may happen at $r=3M$. What is the resolution of this apparent contradiction
?

The answer comes by looking at the region where $3<\alpha$. Once more the function (\ref{qq2}) is one to one and therefore invertible. The domain of the inverse is another copy of the half-line   $[3 \sqrt 3 M,   \infty)$. We denote the corresponding coordinate $\tilde \R$.
The change of variable
\begin{equation}
r =  \frac{\tilde \R}{\sqrt {\alpha(\tilde \R)}}  = \frac{2M\alpha(\tilde \R) } { (\alpha(\tilde \R) -1)}\,. \label{qq2ter}
\end{equation}
maps the half-line $ 3 \sqrt 3 M< \tilde \R < \infty$ onto the interval $3M>r>2M$.  The value $\tilde \R = \infty$ corresponds to the horizon $r=2M$.
The positive value of the acceleration means that test-particles start moving increasing the value of the coordinate $\tilde \R$ and thus they continue their run towards  the horizon. 

\subsection{Inside the horizon}
Let us now consider Equation (\ref{qq2})
\[
 \frac{\R}{M}  = \frac{2 \alpha ^{3/2}} { (\alpha -1)}\,, \label{qq2bis2}
\]
for negative values of $\alpha$; the coordinate $\R$ becomes purely imaginary. By setting $\R \rightarrow i  S$ we see that the  metric
\begin{equation}
ds^2=\frac{4dS^2}{(3-\alpha)^2}-\frac{dt^2}{|\alpha|}-\frac{S^2}{|\alpha|}d\Omega^2\,, \label{sphericmetric3}
\end{equation}
still remains Lorentzian whereas the time and the radial variable have interchanged their roles, as it happens for the Schwarzschild coordinates.
The coordinate  change
\begin{equation}
r = \frac{2M\alpha } { (\alpha -1)} \label{qq2ter2} \, ,  \ \ \mbox{for} \ -\infty< \alpha<0\,,
\end{equation}
now covers the interval $2M>r>0$. Once again
\begin{equation}
\alpha=\frac 1{1-\frac{2 M}{r}}\,, \ \ \mbox{and} \  S =  \frac{r}{\sqrt {\frac{2 M}{r}-1}}\,.
\end{equation}
\subsection{The region inside the horizon is the disformal transformation of a Euclidean space}
To  complete the description of this part of the geometry we observe the following:
consider a metric which is the disformal transformation of the standard four-dimensional Euclidean metric $g_{\mu\nu}$:
\begin{equation}
q_{\mu\nu} = \frac{1}{\alpha} \, g_{\mu\nu} -
\frac{\beta}{\alpha \, (\alpha + \beta) \, w} \, \partial_{\mu} \phi
\, \partial_{\nu} \phi\,,
\label{9junho11bis}
\end{equation}
where $ w=g^{\mu\nu} \, \partial_{\mu} \, \phi \, \partial_{\nu} \phi$.
Let us make the following  choices
\begin{eqnarray}
&& \alpha = -\exp({- 2 \,\phi})\,, \label{potabis}\\[2ex]
&&\alpha+\beta = \frac{(\alpha-3)}{4}^2\,. \label{potbis}
\end{eqnarray}
The difference w.r.t. Eq. (\ref{pota}) is the minus sign at the RHS of Eq. (\ref{potabis}). Let us write the Euclidean metric in spherical coordinates
\begin{equation}
ds^2_E={dt^2}+ d\R^2+ \R^2 d\Omega^2\,.
\end{equation}
\noindent The spherical symmetry together with Eqs.\ (\ref{potabis}) and (\ref{potbis}) imply that the physical metric can be  written as follows:
\begin{eqnarray}
ds^2 &=& \frac{dt^2}{\alpha}+\frac{4d\R^2}{(3-\alpha)^2}+\frac{\R^2}{\alpha}\,d\Omega^2 \nonumber \\[1ex]
&=&  \frac{4d\R^2}{(3-\alpha)^2}- \frac{dt^2}{|\alpha|}- \frac{\R^2}{|\alpha|}\,d\Omega^2\,,  \label{sphericmetricter}
\end{eqnarray}
and is therefore Lorentzian even if the ``background" is Euclidean. As regards the field equation we get:
\begin{equation}
 \Box \phi = 0 \quad \longrightarrow \quad  (3-\alpha)  \partial_\R  \left(\frac{\R^2 (3-\alpha)}{  |\alpha|^\frac{3}2 \alpha} \partial_\R \alpha\right) = 0\,.
\end{equation}
Since we are interested in the region where $\alpha$ is negative we set there $\alpha = -|\alpha| = -\gamma$. The previous equation becomes
 \begin{equation}
   \frac{(3+\gamma)}{  \gamma^\frac{5}2 } \partial_\R \gamma = \frac{4M}{\R^2}\,,
\end{equation}
where $M$ is an integration constant. It follows that
 \begin{equation}
   \frac{(1+\gamma)}{  \gamma^\frac{3}2 }  = \frac{2M}{\R}\,.
\end{equation}
Let us define
\begin{equation}
r =  \frac{\R}{\sqrt {\gamma(\R)}}  = \frac{2M\gamma(\R) } { (1+\gamma(\R))}\,, \label{qq2ter3}
\end{equation}
so that
\begin{equation}
\gamma = - \alpha=- \frac 1{1-\frac{2 M}{r}}\,, \quad \mbox{and } \quad  \R =  \frac{r}{\sqrt {\frac{2 M}{r}-1}}.
\end{equation}
The coordinate change is well-defined in the whole domain of definition of the function $\gamma(\R)$, i.e. the half-line $0 < \R < \infty$, which is mapped onto the interval $0<r<2M$. It is now easy to see that  in that region the line element becomes \begin{equation}
ds^2=\left(1-\frac{2 M}{r}\right){dt^2}-\frac{dr^2}{1-\frac{2 M}{r}}-{r^2}d\Omega^2\,. \label{sphericmetricschwter}
\end{equation}

\section{Other potentials}\label{other_pots}
The precise form of the potential given in Eq. (\ref{13jul2}) which allows for the Schwarzschild solution discussed in the previous section looks however a little artificial and on may wonder what happens in the case of a generic potential, say a power law  $ V =\alpha^{2\n}$.
We renounce with this choice to the existence of the Schwarzschild solution but for certain choices of $\lambda$ it may still be possible to
agree with (part of) the observational regime. Let us explore  again the spherical symmetric solutions.
With the above choice, the  gravitational metric is
\begin{equation}
ds^2 =\frac 1{\alpha} dt^2- \frac{1}{\alpha^{3+2\lambda}} d\R^2- \frac{\R^2}{\alpha}d\Omega^2\,.
\end{equation}
Writing the  field equation in the spherical coordinate system one gets
\begin{equation}
\alpha(\rho)  =\left(1+\frac {A} \R \right)^ {\frac{1}{\n}}\,.
\end{equation}
In the above expression, we have chosen one integration constant equal to 1 so that at infinity the radial coordinate $\R$ coincides with usual spherical coordinate $r$.
Expanding the metric for large values of $\R$ we get
\begin{eqnarray}
ds^2 =\left(1-\frac{A}{\lambda \R}\right) dt^2- \left[ 1-\left(\frac{3}{\lambda}+2\right)\frac{A}{\R}\right]d\R^2 - {\R^2}d\Omega^2\,. \nonumber
\end{eqnarray}
This suggests the following choices
$
A= -4M\,, \lambda = -2\,
$ 
and therefore
\begin{eqnarray}
ds^2 = \sqrt{1-\frac{4M} \R}dt^2 - \frac{d\R^{2}}{\sqrt{1-\frac{4M}{\R}}} - \sqrt{1-\frac{4M}{\R}}{\R^2}d\Omega^2\,.\nonumber
\end{eqnarray}
Here $\rho=4M$ is a true singularity; for instance the scalar curvature
\begin{equation}
R=-{6 M^2 }{\R^{-5/2} (\R-4 M)^{-3/2}}
\end{equation}
diverges there. The potential $ V =\alpha^{-4}$ produces a naked singularity at $\rho=4M$ and the same is true for a generic power-law potential   $ V =\alpha^{2\n}$. We could of course shift the origin and put the singularity in $\R=0$.
Could these models be however useful? We postpone this question to some further investigation.

\section{PPN in GSG}
Let us now discuss the first post-Newtonian approximation of the GSG model admitting the Schwarzschild solution. The field equation in the presence of matter is obtained inserting Eqs. (\ref{pota}),  (\ref{potb}),  and (\ref{13jul2}), into Eq. (\ref{291}) and  in the action (\ref{disf}). We get
\begin{eqnarray}\label{eqorig}
&& \sqrt{\frac{\alpha+\beta}{\alpha^3}}\,\Box\phi
=\left(\frac{\alpha+\beta}{\alpha}\right)^{3/2}\left(\o\Phi +\frac{V'\w}{2V}\right)=\kappa\chi\,, \qquad \label{eqm}
\end{eqnarray}
where
\begin{eqnarray}
&& \chi = \frac{1}{2} \, \left[ \frac{\alpha'}{2\alpha} \, (T-E) +\frac{\alpha'+\beta'}{2(\alpha+\beta)}\,E - C^{\lambda}{}_{;\lambda} \right]\,, \label{xi} \label{chi} \\[2ex]
&&  E=\frac{T^{\mu\nu}\p_{\mu}\phi \p_{\nu}\phi}{(\alpha+\beta)\w} , \quad C^{\lambda}=\frac{\beta}{\alpha}\frac{\left(T^{\lambda\mu}-Eq^{\lambda\mu}\right)}{(\alpha+\beta)\w}\,\p_{\lambda}\phi\,. \quad \label{xi2}
\end{eqnarray}
Some features of this model were already explored in \cite{JCAP} and \cite{gsgcosmo}.  It is more or less obvious from the very beginning that the model cannot fit into the standard Parametrized Post-Newtonian (PPN) formalism  as described in \cite{will}. However we will try to compare this model with that formalism as much as we can.

First of all, the relevant background value to be imposed to the scalar field is $\phi=0$. Then the first term in Eq. (\ref{291}) reduces to the Minkowski metric while the second is undetermined; it is therefore necessary to check that it actually vanishes (at infinity) after solving the approximate field equations.
That said, the post-Newtonian expansion of the scalar field begins with the second order term:
\begin{equation}\label{exp}
\phi \a \phi_{\sc (2)} + \phi_{\sc (4)}\,,
\end{equation}
where $\phi_{\sc (N)}$ denotes the post-Newtonian contribution of order $v^{N}$ (see \cite{will} for an account of the general formalism and the notations).
Let us expand the various ingredients at the respective necessary orders (order 4 at most):
\begin{eqnarray}
&&\alpha  \a 1 -2\phi_{\sc (2)}\,, \quad \beta \a 4\phi_{\sc (2)}\,, \quad \frac{V'}{2V}=\frac{9-\alpha}{3-\alpha} \a 4\,, \\[2ex]
&&\w \a \eta^{\mu\nu}\p_{\mu}\phi_{\sc (2)}\p_{\nu}\phi_{\sc (2)} \a - |\nabla \phi_{(\sc 2)}|^2 =\,\w_{\sc (4)}\,.
\end{eqnarray}
The LHS of Eq. (\ref{eqorig}) becomes
\begin{eqnarray}
&\left(\frac{\alpha+\beta}{\alpha}\right)^{3/2}\left(\o\phi + \frac{V'\w}{2V}\right) \a& -\nabla^2\phi_{\sc (2)} - \nabla^2\phi_{\sc (4)} +4\w_{\sc (4)} \ \nonumber \\[1ex]
&& - 6\phi_{\sc (2)}\nabla^2\phi_{\sc (2)} + \p^2_t\phi_{\sc (2)}\,. \qquad 
\end{eqnarray}
Because of the factor $\kappa=8\pi G/c^{4}$  it is enough to find only the second order expansion of  the RHS of Eq. (\ref{eqorig}). All in all we get:
\begin{eqnarray}
\nabla^2\phi_{\sc (2)} + \nabla^2\phi_{\sc (4)} &+& \ 4 |\nabla \phi_{(\sc 2)}|^2 \ + \ 6\phi_{\sc (2)}\nabla^2\phi_{\sc (2)} - \nonumber \\[1ex]
 &-& \ \p^2_t\phi_{\sc (2)}= -\kappa\chi_{\sc (0)} - \kappa\chi_{\sc (2)}\,.\label{eqfinal}
\end{eqnarray}

\subsection{Second order}
At second order  Eq. \eqref{eqfinal} reduces to
\begin{equation}
\nabla^2\phi_{\sc (2)} = -\kappa\chi_{\sc (0)}\,.
\end{equation}
Let us compute the RHS for a perfect fluid
\begin{equation}
T^{\mu\nu}= \left(\rho+\rho\Pi + p\right)u^\mu u^\nu - pq^{\mu\nu}\,,
\end{equation}
where $u^\mu=(1,v^i)$ is the four velocity of the fluid element and $\Pi$  its specific energy density. Since $\Pi  \a v^2$, the only zeroth order contribution comes from  $T^{00} = \rho$. Eqs. \eqref{xi2} gives
\begin{equation}
T \a \rho\,, \quad E \a {\cal O}(v^2) \quad \mbox{and} \quad C^\lambda_{\ ;\lambda}\a {\cal O}(v^2)\,.
\end{equation}
At  second order , i.e. in the Newtonian limit, Eq. (\ref{eqfinal}) reduces just to Poisson's equation
\begin{equation}\label{eq2}
\nabla^2\phi_{\sc (2)} = \dfrac{\kappa\rho}{2} = 4\pi \rho\,,
\end{equation}
and therefore
 \begin{equation}\label{phi2}
\phi_{\sc (2)}= -\int \frac{\rho(t,x')}{|\vec{x}-\vec{x}'|}\,d^{3}x'\,.
\end{equation}
We complete the analysis by constructing the second order metric  in the monopole approximation.
For regions away from the source we can expand the term appearing in the denominator of \eqref{phi2} in the usual way:
\begin{equation}
\dfrac{1}{|\vec{x}-\vec{x}'|} \a \dfrac{1}{r} + \dfrac{x_i \, x'^i}{r^3}\,,
\end{equation}
where $r=|\vec{x}|$\,. Thus
\begin{equation}\label{h}
\phi_{\sc (2)} \a - \ \frac{M_{\sc (0)}}{r} \ - \ \frac{\vec{x}\cdot \vec{D}_{\sc (0)}}{r^3}\,,
\end{equation}
with
\begin{eqnarray}
M_{\sc (0)} &=& \int{\rho(t,x) \,d^{3}x}\,, \\[2ex]
\vec{D}_{\sc (0)} &=& \int \rho(t,x) \vec{x}\,d^3x\,.
\end{eqnarray}

As usual the dipole term $\vec{D}_{\sc (0)}$ 
can be gauged away by choosing the origin of the  coordinate system at the center of mass of the source (see e.g. (\cite{weinberg})). By doing that we get
\begin{equation}
\phi_{\sc (2)} \a  - \ \frac{M_{\sc (0)}}{r}\,, 
\end{equation}
and  the line element reads
\begin{equation}\label{ds1}
ds^2_{\sc (2)}=\left(1 + 2\phi_{\sc (2)}\right)(dt^2 - r^2d\Omega^{2}) -\left(1 - 2\phi_{\sc (2)}\right)dr^2\,.
\end{equation}
The coordinate transformation \cite{weinberg}
\begin{equation}
r ~ \rightarrow ~R=\left(1+ 2\h \right)r = r + 2M_{\sc (0)}\,,
\end{equation}
puts the line element in the Newtonian form
\begin{equation}
ds^2_{\sc (2)}=\left(1 + 2\phi_{\sc (2)}\right)dt^2 - \left(1 - 2\phi_{\sc (2)}\right)d\vec{X}\cdot d\vec{X}\,,
\end{equation}
where $R^2=\vec{X}\cdot\vec{X}$\,.

\subsection{Fourth order}
The  fourth order  equation is
\begin{equation}\label{eq4}
 \nabla^2\phi_{\sc (4)}+ 4 |\nabla \phi_{(\sc 2)}|^2 + 6\phi_{\sc (2)}\nabla^2\phi_{\sc (2)} - \p^2_t\phi_{\sc (2)}=  - \kappa\chi_{\sc (2)}\,,
\end{equation}
where $\phi_{\sc (2)}$ is the  Newtonian potential given in  Eq.\ (\ref{phi2}).
By introducing the  field
\begin{equation}\label{defpsi}
\psi= \l + 2\h^2\,,
\end{equation}
and using Eq. \eqref{eq2}  we may rewrite the above fourth order equation as follows:
\begin{equation}\label{psi}
\nabla^2\psi= -8\pi\rho\h +\p^2_t\h - \kappa\chi_{\sc (2)}\,.
\end{equation}
To obtain an explicit expression for $\chi_{\sc (2)}$ we note that  at second order
\begin{equation}
q_{00} \a 1+2\h, \quad (u^0)^2 \a 1 + v^2 - 2\h\, ,
\end{equation}
and the components of the energy momentum tensor and its trace are given by
\begin{eqnarray}
 T^{00} &\a& \rho +\rho\Pi + \rho v^2 - 2\rho\h\,, \\[2ex]
 T^{0i}  &\a& \rho v^i\,,\\[2ex]
 T^{ij}  &\a& \rho v^i v^j + p\delta^{ij}\,,
\end{eqnarray}
\begin{equation}
T\a T^{00}(1+2\h) + T^{ij}\eta_{ij} \a \rho + \rho\Pi -3p\,.
\end{equation}
The other terms appearing in $\chi$ are
\begin{eqnarray}
E &=& \frac{1}{(\alpha+\beta)\w}\,T^{\mu\nu}\p_\mu\phi \p_\nu\phi \nonumber \\[1ex]
&\a& \ -\frac{ \left(D_t \h\right)^2}{ |\nabla \phi_{(\sc 2)}|^2} \rho- p \,,\label{a}\\[2ex]
C^{\lambda} &=& \dfrac{\beta}{\alpha(\alpha+\beta)}\,\dfrac{\left(T^{\lambda\mu} - E q^{\lambda\mu}\right)}{\w}\,\p_\mu\phi \nonumber \\
&\a& \ -\frac{4\h}{ |\nabla \phi_{(\sc 2)}|^2}\left(T^{\lambda\mu}-Eq^{\lambda\mu}\right)\,\p_\mu\h\,. \label{b}
\end{eqnarray}
where
\be
D_t \h= \p_t\h + v^i\p_i\h.
\ee
Because of the divergence of $C^{\lambda}$ in Eq. (\ref{chi}) the temporal component $C^0$ is needed only at first order:
\begin{equation}
C^0_{\sc (1)} = -\dfrac{2\rho\, D_t \h^2}{ |\nabla \phi_{(\sc 2)}|^2}\,.
\end{equation}
As for the spatial components we have
\begin{equation}
C^i_{\sc (2)} = -\frac{2 D_t \h^2}{ |\nabla \phi_{(\sc 2)}|^2}\, \rho v^i + \frac{4\rho\h (D_t \h)^2 }{ |\nabla \phi_{(\sc 2)}|^4}\delta^{ij}\p_j\h\,.
\end{equation}
Inserting the above expressions in Eq. \eqref{xi} and collecting only the relevant terms we get
\begin{eqnarray}
\chi_{\sc (2)} &=& -\frac{1}{2}\Bigg[\rho\Pi -p + \dfrac{2\rho (D_t \h)^2 }{ |\nabla \phi_{(\sc 2)}|^2} + \p_t C^0_{\sc (1)} + \nonumber\\[1ex]
&& + \ \p_i C^i_{\sc (2)}\Bigg].
\end{eqnarray}
Finally, the fourth order equation \eqref{psi} is given by
\begin{eqnarray}
\nabla^2 \psi &=& - \ 8\pi\rho\h +\p^2_t\h +4\pi \Bigg[\rho\Pi -p + \nonumber \\[1ex]
&& + \ \frac{2\rho (D_t \h)^2}{|\nabla\phi_{(\sc 2)}|^2} + \p_t C^0_{\sc (1)} + \p_i C^i_{\sc (2)} \Bigg] \,. \label{4}
\end{eqnarray}

\subsection{Static case}
While it is easy to integrate Eq.\ (\ref{4}) by using the fundamental solution of the Laplace operator,
the so-obtained solution is not amenable to a classification in terms of the standard ten PPN potentials of Will and Nordvedt \cite{will}.
This might not come as a surprise: the standard PPN formalism is also not enough to describe the post-Newtonian limit say scalar-tensor theory of the Horndeski class \cite{horn}.

In principle it may be conceived that an infinite number of parameters is necessary to parametrize any metric theory of gravity \cite{ciufolini}. In the Will-Nordvedt approach the post-Newtonian metric is however parameterized in the standard gauge in terms of ten numerical coefficients as follows:
\begin{eqnarray}
 g_{00} &=& 1-2U+ 2\beta U^2 - (2 \gamma +2+\alpha_3 +\zeta _1-2 \xi ) \Phi_1 - \nonumber \\[1ex]
 && - \ 2(3 \gamma -2\beta+1+\zeta _2+ \xi ) \Phi_2 -2(1+\zeta _3 ) \Phi_3 - \nonumber \\[1ex]
 && - 2(3 \gamma +3\zeta _4-2 \xi ) \Phi_4 + (\zeta _1-2 \xi ) {\cal A}  +2\xi \Phi_W, \qquad \\[2ex]
g_{0i} &=&  \frac 1 2(4 \gamma +3+\alpha_1-\alpha_2+ \zeta_1-2\xi) V_i +\nonumber \\[1ex]
 &&+ \ \frac 1 2(1+\alpha_2- \zeta_1+2\xi) W_i\,,\\[2ex]
g_{ik} &=& -(1+2\gamma U) \delta_{ik}\,.
\end{eqnarray}
where $\bh=-U$. The fluid velocity enters in the integrals expressing the potentials  $\Phi_1$,  $V_i$, $W_i$ (i=1,2,3), $\cal A$.
One may also observe that corresponding coefficients are all written in terms of $(\gamma, \xi)$ and $( \alpha_1,\alpha_2,\alpha_3, \zeta _1)$.
The potentials  $\Phi_W$, $\Phi_2$,  $\Phi_3$, $\Phi_4$ do not refer to the fluid velocity.
The corresponding coefficients are all written in terms of $(\gamma, \xi,)$ and $(\beta, \zeta_2,\zeta_3, \zeta _4)$. 
Given the values of $\gamma$ and $\xi$ the two subsets of coefficients $( \alpha_1,\alpha_2,\alpha_3, \zeta _1)$ and $(\beta, \zeta_2,\zeta_3, \zeta _4)$ are determined independently. In particular for theories that fit well in the standard PPN formalism, the coefficients $(\beta, \zeta_2,\zeta_3, \zeta _4)$ may be determined in the static approximation where the potentials $\Phi_1\,, \ {\cal A}\,, \ V_i $\,, and $W_i$ vanish.

If we limit ourselves to the static approximation 
Eq. \eqref{4} simplifies to
\begin{eqnarray}
\nabla^2 \psi= -8\pi\rho\h + 4\pi\rho\Pi - 4\pi p \,.
\end{eqnarray}
In terms of the standard PPN potentials
\begin{eqnarray}
\nabla^2\Phi_2 &=& 4\pi\rho\h\,,\\[2ex]
\nabla^2\Phi_3 &=& -4\pi\rho\Pi\,, \\[2ex] 
\nabla^2\Phi_4 &=& - 4\pi p\,,
\end{eqnarray}
we get
\begin{equation}
\psi= -2\Phi_2 - \Phi_3 + \Phi_4\,.
\end{equation}
In constructing the metric we use again the monopole approximation; the potentials $\phi_{\sc (2)}\,, \ \Phi_2\,, \ \Phi_3$\, and $\Phi_4$ depend only on $r$. The spatial components of the metric are needed at second order and we already computed them;
$q_{00}$ is needed at fourth order:
\begin{eqnarray}
q_{00}
&\a &1 + 2(\h + \l) + 2\h^2\,, \nonumber \\[1ex]
&= &1 + 2\h - 2\h^2 - 4\Phi_2 - 2 \Phi_3 + 2\Phi_4\,.
\end{eqnarray}
The line element is as in Eq. \eqref{ds1} with the $q_{00}$ appearing there replaced by the one given above.
Once again we need to perform a coordinate transformation to write the spatial part of the metric in the standard PPN gauge.
But the spatial part did not change and the transformation will be the same:
\begin{equation}
r ~ \rightarrow ~R=\left(1+ 2\h \right)r = r + 2M_{\sc (0)}\,.
\end{equation}
Now  $\h$ is rewritten as follows
\begin{equation}
\h =\frac{M_{\sc (0)}}{r} = \frac{M_{\sc (0)}}{R}\left(1- 2\,\frac{M_{\sc (0)}}{R}\right)= \bh - 2\bh^2\,.
\end{equation}
Finally the line element in the static monopolar approximation is given by
\begin{eqnarray}
ds^2_{\sc (2)} &=& \left(1+ 2\bh +2\bh^2 - 4\Phi_2 - 2\Phi_3 + 2\Phi_4 \right)dt^2 - \nonumber \\[1ex]
&& -  \ \left(1 - 2\bh\right)d\vec{X}\cdot d\vec{X}\,, \label{line}
\end{eqnarray}
where $R^2=\vec{X}\cdot\vec{X}$\,. With all the above simplifications we get
\begin{align}
 \alpha=1, \ \  \beta=1, \ \ \gamma=1,\ \ \xi=0,\\[1ex]
 \zeta_2=0,\ \  \zeta_3=0,\ \  \zeta_4=-\frac{4}{3}\,.
\end{align}

The first observation is  that the parameter $\gamma$ has the right value 1 while in any Nordstr\"om model one has $\gamma=-1$.
On the other hand the parameter  $\zeta_4=-\frac{4}{3}$ seems to indicate a violation of energy-momentum conservation. But the proof of this statement  \cite{will} crucially relies on the assumptions behind the standard PPN formalism and does not seem to apply in the present case.
Indeed, in the model we are describing here there is at least one obvious conserved quantity  that may be related to the stress-energy content of gravity.
That is the  Einstein frame energy-momentum tensor which is locally conserved in the standard sense for any scalar theory of the class (\ref{disf}).

\section{Some concluding remarks and outlook}

In this paper we have introduced a family of geometric models for gravity based on a scalar field 
that (include and) generalize the Nordstr\"om class of scalar theories. The important point is that the geometrical prior is now a disformal transformation of a fiducial metric (here the Minkowski metric) rather than a conformal transformation as in the Nordstr\"om case.
Extensions of General Relativity obtained by coupling the matter to a disformally transformed  metric 
have been shown by Bekenstein \cite{beke} to preserve  causality and the weak equivalence principle. 
Here we  propose that it may be interesting to take a huge step backward 
and explore models in which the gravitational degrees of freedom are just described by the field $\phi$ .

The class of models introduced in this way includes in particular a model 
that we have studied in two preceding papers (GSG)
One of the criteria used to select this model \cite{JCAP}
was the requirement that it should admit the Schwarzschild solution.
We have deepened the discussion about this point 
and see that the disformal hypothesis is more subtle than one may think at first sight. 
In particular we have seen that three charts are necessary to cover the Schwarzschild solution and that the region inside the horizon is somehow surprisingly the disformal transformation of a Euclidean metric.

Next we have discussed  the first post-Newtonian limit of GSG. The theory does not fit well within the standard PPN formalism. Even so, considering a static approximation, is possible to read some of the PPN parameters which evidence the improvement brought by GSG in the realm of scalar theories of gravitation.

In the end, GSG's based on one scalar field present a few interesting features which
capture some aspects of gravity. The move towards more realistic models would consist in enriching the theory by adding a second
scalar field. This is because there are two quantities that one can associate to a gravitational field: mass and angular momentum. 
It is more or less obvious that the form
used in GSG does not allow for rotating geometries that should correspond to a physical angular momentum not to mention more general gravitational configurations. We will discuss
this GSG extension elsewhere.

\begin{acknowledgments}
We thank Nathalie Deruelle for several illuminating discussions and for a correspondence. J.D.T. was supported by CNPq (Brazil), process 161122/2014-0. M N. acknowledges a grant from CNPq.
\end{acknowledgments}

\end{sloppypar}
\end{document}